# Quantum Error Correction
Todd A. Brun

## Summary

**Quantum error correction** is a set of methods to protect quantum information—that is, quantum states—from unwanted environmental interactions (*decoherence*) and other forms of noise. The information is stored in a *quantum error-correcting code*, which is a subspace in a larger Hilbert space. This code is designed so that the most common errors move the state into an *error space* orthogonal to the original code space while preserving the information in the state. It is possible to determine whether an error has occurred by a suitable measurement and to apply a unitary correction that returns the state to the code space, without measuring (and hence disturbing) the protected state itself. In general, codewords of a quantum code are entangled states. No code that stores information can protect against all possible errors; instead, codes are designed to correct a specific *error set*, which should be chosen to match the most likely types of noise. An error set is represented by a set of operators that can multiply the codeword state.

Most work on quantum error correction has focused on systems of quantum bits, or *qubits*, which are two-level quantum systems. These can be physically realized by the states of a spin-1/2 particle, the polarization of a single photon, two distinguished levels of a trapped atom or ion, the current states of a microscopic superconducting loop, or many other physical systems. The most widely-used codes are the *stabilizer codes*, which are closely related to classical *linear* codes. The code space is the joint +1 eigenspace of a set of commuting Pauli operators on $n$ qubits, called *stabilizer generators*; the *error syndrome* is determined by measuring these operators, which allows errors to be diagnosed and corrected. A stabilizer code is characterized by three parameters $[[n, k, d]]$, where $n$ is the number of *physical* qubits, $k$ is the number of encoded *logical* qubits, and $d$ is the minimum distance of the code (the smallest number of simultaneous qubit errors that can transform one valid codeword into another). Every useful code has $n > k$; this physical redundancy is necessary to detect and correct errors without disturbing the logical state.

Quantum error correction is used to protect information in *quantum communication* (where quantum states pass through noisy channels) and *quantum computation* (where quantum states are transformed through a sequence of imperfect computational steps in the presence of environmental decoherence to solve a computational problem). In quantum computation, error correction is just one component of *fault-tolerant* design. Other approaches to error mitigation in quantum systems include *decoherence-free subspaces*, *noiseless subsystems*, and *dynamical decoupling*.

**Introduction**

Quantum error correction (QEC) is a set of techniques to protect quantum states from the effects of environmental noise, or decoherence (Gaitan, 2008; Lidar and Brun, 2013; Nielsen and Chuang, 2000; Suter and Alvarez, 2016).  In less than twenty-five years the subject has gone from a time when many prominent quantum theorists doubted that QEC was even possible to a large field with a well-developed theory, thousands of published papers, and international conferences.  It has been a remarkable trajectory, which has both paralleled and made possible the trajectory of the broader field of quantum information science.  Without quantum error correction, quantum computers would be restricted to problems so small that they would be of little use or interest.  With quantum error correction, there are strong theoretical reasons to believe that quantum computations of any size can be done without requiring vast improvements in technology.  Indeed, starting around 2014 and accelerating since then, companies have begun competing with each other to build the first small, noisy quantum computers, hoping to pave the way towards the powerful quantum technologies of tomorrow.

QEC is built on the theory of quantum codes, which was first developed in the mid-1990s, and which has been greatly expanded and elaborated since then.  A standard quantum code stores quantum information—a quantum state in a (usually) finite-dimensional Hilbert space—in a subspace of a higher-dimensional Hilbert space.  For a code to protect a quantum state against a set of errors, that subspace must be chosen so that each error transforms the state in such a way that it is possible to deduce which error occurred (by performing a suitable measurement) without acquiring any information about the state that was stored in the code, which would necessarily disturb it.  It is remarkable that this is possible at all; but as we shall see, it is not only possible, but can be done both robustly and efficiently.

This article will briefly introduce the most important aspects of quantum error correction.  Given the size and diversity of the field, it is impossible to describe every important idea in an article of this length, let alone the technical details.   But it should serve as a starting point for anyone who wants to understand a key building block of future quantum information technology.

**A Brief History of Quantum Error Correction**

The idea of QEC was driven by the explosion of interest in quantum computers.  A handful of people, starting in the 1980s, began to ask if a computer operating according to the laws of quantum mechanics might be more powerful than ordinary classical computers, which obey classical laws.  A slow trickle of results (Benioff, 1980; Deutsch, 1985; Deutsch, 1989; Deutsch and Josza, 1992; Feynman, 1982; Manin, 1980; Simon, 1994; Yao, 1993) showed that there were certain problems where a quantum computer could outperform a classical computer.  But widespread interest was created in 1994 when Peter Shor (then at AT&T Research) proved that a quantum computer could factor large integers into their prime factors efficiently (Shor, 1994).  The computational difficulty of factoring underlies common public-key encryption systems (like RSA) that guarantee the security of Internet transactions, so Shor's result sparked both interest and concern.

It also sparked widespread skepticism (Unruh, 1995; Landauer, 1996). Shor's algorithm assumed that the quantum computer evolved ideally, in perfect isolation from its environment. The early 1990s also saw the general realization that large quantum systems are almost impossible to isolate, producing environmental *decoherence* that would cause errors and destroy the quantum evolution required for the algorithm.

Of course, classical computers are also subject to errors, and there is a well-known cure: *error-correcting codes*. The simplest example of such a code is the *repetition* code (or *majority rule* code), in which a single bit value 0 or 1 is duplicated multiple times: $0 \to 000$ and $1 \to 111$. If an error flips one of the bits, it can be detected by comparing the values of all three bits, and corrected by taking a majority vote. The question was raised whether quantum computers could also be protected by error-correcting codes.

Naïvely, it seemed like this should not be possible. A quantum state $|\psi\rangle$ cannot be copied redundantly (i.e., $|\psi\rangle \otimes |\psi\rangle \otimes |\psi\rangle$) because of the *no-cloning theorem* (Dieks, 1982; Park, 1970; Wootters and Zurek, 1982). Measuring the system, so it could be copied classically, would disturb the state, just as decoherence does. Two prominent quantum experimenters, Serge Haroche and Michel Raimond, published an article in *Physics Today* entitled "Quantum Computing: Dream or Nightmare?" (Haroche and Raimond, 1996) in which they cast doubt on the practicality of quantum computers. (Serge Haroche later won the Nobel Prize for his groundbreaking work bringing quantum computers closer to reality.)

Fortunately, the naïve perspective was not the final word. In 1995, Peter Shor published a paper (Shor, 1995) in which he demonstrated a 9-qubit quantum error-correcting code (QECC) that could correct an arbitrary error on any single qubit. Rather than *copying* the state, the encoding spreads the quantum state nonlocally over all nine of the qubits in an entangled state, in such a way that local errors do not irreparably destroy the stored information. Working independently, Andrew Steane published a 7-qubit code in 1996 (Steane, 1996a) with a structure based on a classical linear code, the Hamming code. Laflamme, Miquel, Paz and Zurek, and independently Bennett, DiVincenzo, Smolin and Wootters (Bennett et al., 1996; Laflamme et al., 1996) found a 5-qubit code in 1996 and proved that this is the shortest possible code that can correct a general quantum error. Knill and Laflamme, and independently Bennett, DiVincenzo, Smolin and Wootters also found a criterion for when a QECC could correct a given set of errors (Bennett et al., 1996; Ekert and Macchiavello, 1996; Knill and Laflamme, 1997).

These results only proved that quantum information could be stored with protection against errors; for a quantum computer, that information would also have to be processed. The way for this was paved in a series of papers (Aharonov, 1999; Aharonov et al., 1996; Aharonov and Ben-Or, 1997; Aharonov and Ben-Or, 1999; Aliferis, Gottesman and Preskill, 2006; DiVincenzo and Shor, 1996; Gottesman, 1998; Kitaev, 1997a; Kitaev, 1997b; Knill, Laflamme and Zurek, 1998a; Knill, Laflamme and Zurek, 1998b; Preskill, 1997; Preskill, 1998a; Preskill, 1998b; Reichardt, 2005a; Shor 1996) showing that quantum computation can be done *fault-tolerantly*: that is, that protection against errors can be maintained during the processing of quantum information, and even during the error-correction process itself. This work did not convince all the skeptics at once. Indeed, there are still a handful of prominent quantum-computing skeptics (Dyakonov, 2019; Kalai, 2011; Moskvitch, 2018). But the theory of quantum error correction has become ever more powerful and convincing over time, and experimental tests have given cause for optimism. As of this writing (in 2019), essentially everyone in the field understands both the capabilities and the requirements of quantum error correction, and large-scale efforts are underway to realize them.

**Decoherence and Quantum Noise**

*Environmental decoherence.* The Schrödinger equation

$$i\hbar \frac{d|\psi\rangle}{dt} = H|\psi\rangle,$$

describes the evolution of quantum systems in isolation, where $|\psi\rangle$ is the state vector (written in Dirac notation). These closed systems have a well-defined Hamiltonian operator $H$, which gives complete information about how these systems evolve. The resulting evolution is *unitary*: the evolution of the state is given by a linear map $|\psi(0)\rangle \rightarrow |\psi(t)\rangle = U|\psi(0)\rangle$ where $U^\dagger U = UU^\dagger = I$. Note that these Hamiltonians may "come from outside" the system; for instance, we can turn external fields on and off, shine lasers, etc. What makes a quantum system closed is that it doesn't act *back* on the external world. The external fields, lasers, etc., can all be treated as classical potentials.

The unfortunate reality is that this idealization is a fiction. All real quantum systems interact with the outside world at least weakly; and the existence of interactions which allow us to manipulate a system (as needed for quantum information processing) also allows the system to interact with the external environment. This environmental interaction is called *decoherence*.

Two things happen in decoherence. First, random influences from the outside can perturb the system's evolution, as if some random Hamiltonian was turned on, in addition to the usual Hamiltonian. Second, the interaction between the system and environment can cause information about the system to leak into the environment. This information leakage leaves the system correlated with the environment. The effect on the system is as if unwanted measurements have been performed (without, in general, our knowing the measurement results).

In fact, these two processes generally both occur, and the practical effects of them often look similar. Indeed, in quantum mechanics there is no sharp distinction between them. If decoherence persists long enough, it is possible for all information about the original state of the system to be lost. In the shorter term, decoherence can destroy quantum effects such as interference and entanglement (on which quantum information processing depends). Indeed, decoherence is the main reason why quantum effects are not perceived at familiar classical scales; any large-scale superposition (like an alive-and-dead cat) would decohere almost instantaneously, leaving a mixed state that acts just like a classical probability distribution.

*CPTP maps.* In quantum information processing, time evolution is usually treated as discrete, representing the total evolution over a finite time interval (e.g., one computational step). Decoherence is therefore treated as a discrete-time map. In general, we must describe the state of a quantum system in terms of a density matrix $\rho$ rather than a state vector $|\psi\rangle$, to allow for the possibility that the state is *mixed* (i.e., contains uncertainties that represent missing information) rather than *pure* (isolated and perfectly known). A pure state $|\psi\rangle$ has a density operator description $\rho = |\psi\rangle\langle\psi|$, which is a rank-1 projector. A density matrix is a positive Hermitian operator with trace equal to 1 (representing the total probability). Maps that represent decoherence must preserve these properties: they are *completely positive, trace-preserving* (CPTP) maps. CPTP maps can be written in the form

$$\rho \to \rho' = \sum_k A_k \rho A_k^\dagger, \qquad \sum_k A_k^\dagger A_k = I,$$

where the operators $\{A_k\}$ are called *Kraus operators*.

In general, the Kraus decomposition of a CPTP map is not unique, but one can approximately think of the map as the state $|\psi\rangle$ being multiplied by one of the operators $A_k$ chosen at random with probability $p_k = \langle\psi|A_k^\dagger A_k|\psi\rangle$. Since one does not know which operator has multiplied the state, one uses a mixture of all of them.

*Random unitaries.* In a similar way, if an unknown influence is applied to the quantum system from the outside, we can model that as a set of unitaries $\{U_k\}$ that occur with respective probabilities $\{p_k\}$. Here, again, one would describe the state of the system as a mixture of all possible evolved states:

$$\rho \to \rho' = \sum_k p_k U_k \rho U_k^\dagger, \qquad \sum_k p_k = 1.$$

In this case again we have a CPTP map, and we can define the Kraus operators to be $A_k \equiv \sqrt{p_k} U_k$. Note that the randomness in the unitary evolution need not be due to outside influence: it could also be from uncertainty of the Hamiltonian, due to imperfect control of the system or any other reason. CPTP maps give a unified description of all possible sources of *Markovian* (i.e., time-local) noise, and in quantum information science one does not usually make a sharp distinction between different noise sources.

**Qubits and Pauli Operators**

*Qubits.* The canonical quantum system used in quantum computation and quantum information is the *quantum bit* or *qubit* (Schumacher, 1995). This is a system with two distinct levels, whose state is in a two-dimensional complex Hilbert space $\mathcal{H} = \mathbb{C}_2$. Examples of such systems are the spin of a spin-1/2 particle (like an electron whose spin can be up or down along an axis in space) or the polarization of a single photon (which can be horizontally or vertically polarized). We choose a standard basis $\{|0\rangle, |1\rangle\}$ for the Hilbert space of a single qubit (often called the *computational basis*), which is orthonormal: $\langle i|j\rangle = \delta_{ij}$, $i,j = 0,1$. The standard basis is also often called the *Z basis*, because it is the eigenbasis of the Pauli Z operator (see below). $\{|0\rangle, |1\rangle\}$ represent column vectors:

$$|0\rangle = \begin{pmatrix} 1 \\ 0 \end{pmatrix}, \qquad |1\rangle = \begin{pmatrix} 0 \\ 1 \end{pmatrix}.$$

*The Pauli operators.* We can write any operator $O$ on $\mathbb{C}_2$ as a $2 \times 2$ complex matrix. Any such matrix $O$ can be written as a linear combination of the identity matrix $I$ and the three *Pauli operators*, $X, Y, Z$:

$$I = \begin{pmatrix} 1 & 0 \\ 0 & 1 \end{pmatrix}, \qquad X = \begin{pmatrix} 0 & 1 \\ 1 & 0 \end{pmatrix}, \qquad Y = \begin{pmatrix} 0 & -i \\ i & 0 \end{pmatrix}, \qquad Z = \begin{pmatrix} 1 & 0 \\ 0 & -1 \end{pmatrix}.$$

The Pauli operators — first introduced to describe the algebra of spin-1/2 particles — have interesting algebraic properties. They are Hermitian, unitary, and traceless, with eigenvalues $\pm 1$. They mutually anticommute, and generate a closed group:

$$X^2 = Y^2 = Z^2 = I,$$
$$XY = -YX = iZ,$$
$$YZ = -ZY = iX,$$
$$ZX = -XZ = iY.$$

*Quantum registers and the Pauli Group.* While the mathematics of a single qubit is surprisingly rich, it is still very limited in its use: there is not much information processing that can be done with a single quantum bit. More generally one has a collection of $n$ qubits, called a *quantum register* or a *quantum codeword*. This joint system has an associated Hilbert space $\mathcal{H}^{\otimes n} = \mathbb{C}_2 \otimes \mathbb{C}_2 \otimes \cdots \otimes \mathbb{C}_2$, the $n$-fold tensor product of the single-qubit space, which has dimension $2^n$. We can identify a standard basis for the quantum register:

$$|i_1 i_2 \ldots i_n\rangle \equiv |i_1\rangle \otimes |i_2\rangle \otimes \cdots \otimes |i_n\rangle, \qquad i_j \in \{0,1\} \; \forall j.$$

In quantum error correction we often need to consider multiplying these basis states for $n$ qubits by tensor products of Pauli operators. To do this it is convenient to define the *Pauli Group on $n$ qubits* as the set of all $n$-fold tensor products of Pauli operators and the identity: $\mathcal{G}_n = \{i^l O_1 \otimes O_2 \otimes \cdots \otimes O_n\}$, where $i = \sqrt{-1}$, $l = 0,1,2,3$, and $O_j \in \{I, X, Y, Z\} \; \forall j$. It is not hard to see that this set of operators is closed under multiplication and forms a group. Every operator in this group has eigenvalues either $\pm 1$ or $\pm i$. For compactness we will often omit the tensor-product symbol $\otimes$ when the meaning is clear. For instance, for $n = 3$, we can write $X \otimes I \otimes X \equiv XIX$, $Z \otimes Z \otimes Y \equiv ZZY$, and so forth. We will use this notation for the rest of this article.

**Error Models and Simple Quantum Error-Correcting Codes**

*Error models.* Before defining the notion of a quantum error-correcting code (QECC), we should first establish what we mean by an *error* or an *error model*. Recall that quantum systems subject to decoherence and other sources of noise evolve by CPTP maps, and that these can be thought of as a set of Kraus operators which multiply the state of the system with some probabilities. We define an *error set* $\mathcal{E} = \{E_j\}$ as a set of operators proportional to Kraus operators. Generally, at least one of these operators (usually $E_0$) is taken to be the identity $I$ (at least to a good approximation), which corresponds to no error occurring, while the others represent possible errors.

*The bit-flip code.* Let us restrict ourselves for the present to systems of qubits; and let us further assume that these qubits undergo *independent, identically distributed noise* (i.i.d. noise), meaning that the CPTP map acting on the quantum register is the product of identical CPTP maps acting on the individual qubits. We can then define the simplest possible QECC based on the example of the classical repetition code. Suppose that each of the qubits is independently subject to *bit-flip* noise:

$$\rho \to \rho' = (1-p)\rho + pX\rho X.$$

The Pauli operator $X$ acts as a bit flip, because $X|0\rangle = |1\rangle$ and $X|1\rangle = |0\rangle$, and $p$ is the probability of a bit flip per timestep. We protect the qubit state $|\psi\rangle = \alpha|0\rangle + \beta|1\rangle$ by encoding it as a 3-qubit *codeword* $|\psi_L\rangle = \alpha|000\rangle + \beta|111\rangle$. The set of all possible codewords forms a 2-dimensional subspace of the 8-dimensional Hilbert space $\mathcal{H}^{\otimes 3} = \mathbb{C}_2 \otimes \mathbb{C}_2 \otimes \mathbb{C}_2$. This subspace is called the *code space*. (Note that most states in this space are highly entangled.)

All three of these qubits are subject to identical bit-flip noise. This error model has the error set $\mathcal{E} = \{I, X_1, X_2, X_3, X_1X_2, X_1X_3, X_2X_3, X_1X_2X_3\}$ where $X_j$ denotes the Pauli $X$ acting on qubit $j$:

$$X_1 \equiv XII, \qquad X_2 \equiv IXI, \qquad X_3 \equiv IIX.$$

For this error model, the weight-1 errors ($X_1, X_2, X_3$) all have probability $p(1-p)^2$, the weight-2 errors ($X_1X_2, X_1X_3, X_2X_3$) have probability $p^2(1-p)$, the weight-3 error ($X_1X_2X_3$) has probability $p^3$, and the weight-0 (identity) error has probability $(1-p)^3$.

Classically, single bit-flip errors are corrected by measuring the three bits of the codeword and taking a majority vote. Clearly this would violate the purpose of a QECC: measuring the bits would project the system into one of the basis states with probability $|\alpha|^2$ or $|\beta|^2$, destroying the superposition state that we are trying to protect. So how can error correction be done? Note what happens to the codeword state under the identity and the three weight-1 errors:

$$|\psi_L\rangle = \alpha|000\rangle + \beta|111\rangle \xrightarrow{I} \alpha|000\rangle + \beta|111\rangle,$$
$$|\psi_L\rangle = \alpha|000\rangle + \beta|111\rangle \xrightarrow{X_1} \alpha|100\rangle + \beta|011\rangle,$$
$$|\psi_L\rangle = \alpha|000\rangle + \beta|111\rangle \xrightarrow{X_2} \alpha|010\rangle + \beta|101\rangle,$$
$$|\psi_L\rangle = \alpha|000\rangle + \beta|111\rangle \xrightarrow{X_3} \alpha|001\rangle + \beta|110\rangle.$$

All four of these states are mutually orthogonal for all values of $\alpha, \beta$. They lie in orthogonal subspaces. Therefore, there is a quantum measurement that will tell which of these four subspaces the state is in *without* projecting onto a basis state and thereby destroying the superposition. Once one knows which subspace the state is in, it is possible to transform the state back to $|\psi_L\rangle$ by applying one of the operators $I, X_1, X_2, X_3$, which are all unitary; applying this operator also does not require us to know what state is being stored. This is the key insight that makes quantum error correction possible: for a properly designed QECC, there is a measurement that reveals the error without revealing any information about the encoded state.

For this particular code, it is not hard to see that measuring which subspace the state is in is equivalent to measuring the two commuting observables $Z_1 Z_2 \equiv ZZI$ and $Z_2 Z_3 \equiv IZZ$. The four states above are eigenstates of both of these observables with eigenvalues ±1. Measuring $Z_1 Z_2$ (or $Z_2 Z_3$) is equivalent to measuring the parity of qubits 1 and 2 (or 2 and 3); moreover these are *joint* observable on two qubits, which can be measured without measuring the values of Z on the individual qubits. These observables $Z_1 Z_2$ and $Z_2 Z_3$ are called *stabilizer generators* of this code, and the values of these two observables give the *error syndrome* that identifies the error that occurred. For this code that would be one of the four outcomes (+1,+1), (+1,-1), (-1,+1), (-1,-1). (More details can be found in the subsection on Stabilizer groups and their generators, and in the caption of Fig. 2.)

This bit-flip code has a *correctable error set* with four error operators: $\{I, X_1, X_2, X_3\}$. However, the full error set of this error model contained *eight* error operators. The three errors of weight-2 and one error of weight-3 are *uncorrectable errors*. They produce states in the same four subspaces above, and it is easy to see that in the case of those high-weight errors the correction procedure will produce the erroneous state $|\psi_L'\rangle = \alpha|111\rangle + \beta|000\rangle$. In fact, the weight-3 error will not even be recognized as an error: it is an *undetectable error*. This is a general property of QECCs: *no QECC can correct every possible error*. (This is also true of classical error-correcting codes.) In practice, the goal is to choose a code that can correct the *most likely* errors. For the error model discussed here, the probability of success is the probability of either no error or a weight-1 error: $(1-p)^3 + 3p(1-p)^2$. The probability of failure is $3p^2(1-p) + p^3$. If the original single qubit state $|\psi\rangle = \alpha|0\rangle + \beta|1\rangle$ had been left unencoded, it would have probability of success $1-p$ and probability of failure $p$. So this code gives an improved success probability if $3p^2(1-p) + p^3 < p$, corresponding to an error probability per qubit of $p < 1/2$.

*The phase-flip code.* For a classical bit channel, the only errors that are possible are flipping a bit from 0 to 1 or vice versa. That is decidedly *not* the case for qubit channels, where any 2-dimensional operator E could constitute an error. One particular type of error, that has no classical equivalent, is a *phase-flip error*: $|\psi\rangle = \alpha|0\rangle + \beta|1\rangle \rightarrow Z|\psi\rangle = \alpha|0\rangle - \beta|1\rangle$, which applies a relative phase factor of -1 between the two basis states. The code we designed against bit flips is useless against this type of error: $Z_j(\alpha|000\rangle + \beta|111\rangle) \rightarrow \alpha|000\rangle - \beta|111\rangle$ for any phase-flip error $Z_j, j = 1,2,3$. To protect against Z errors, we can use a code expressed in the X basis, rather than the standard Z basis:

$$|\pm\rangle = \frac{1}{\sqrt{2}}(|0\rangle \pm |1\rangle), \quad X|\pm\rangle = \pm|\pm\rangle, \quad Z|\pm\rangle = |\mp\rangle.$$

So in this basis, Z acts like a bit flip. In terms of these basis states, we can encode the state $|\psi\rangle = \alpha|0\rangle + \beta|1\rangle$ as $|\psi_L\rangle = \alpha|+++\rangle + \beta|---\rangle$. Using this code to detect and correct phase-flip errors without disturbing the encoded state works exactly like the bit-flip code, but with the basis change $|0\rangle \rightarrow |+\rangle, |1\rangle \rightarrow |-\rangle$. The error syndrome for this code is determined by measuring the two stabilizer generators $X_1 X_2 \equiv XXI$ and $X_2 X_3 \equiv IXX$.

While the bit-flip and phase-flip codes prove that it is possible to design a QECC capable of correcting a finite set of errors, this seems like a very limited result.  Each of them is essentially the classical repetition code in a particular choice of basis; they can correct their own set of errors but are useless against each other's.  Moreover, the set of all possible quantum errors forms a continuum, since any linear operator $E$ could, in principle, be an error.  On the face of it, it might seem like quantum error correction is too limited to provide real protection to quantum states in the presence of realistic decoherence.  It turns out, however, that QECCs can be far more powerful than these examples might suggest.

*The Shor code.*  In 1995, Peter Shor published a 9-qubit QECC that was capable of correcting *any arbitrary error* on a single qubit, and could protect one logical qubit state (Shor, 1995).  A qubit state $|\psi\rangle = \alpha|0\rangle + \beta|1\rangle$ was encoded as $|\psi_L\rangle = \alpha|0_L\rangle + \beta|1_L\rangle$.  The code-space basis is

$$|0_L, 1_L\rangle = \frac{1}{2\sqrt{2}}(|000\rangle \pm |111\rangle)\otimes(|000\rangle \pm |111\rangle)\otimes(|000\rangle \pm |111\rangle),$$

where the + sign goes with the state $|0_L\rangle$ and the – sign with $|1_L\rangle$.  A little unpacking is needed to see how this code works.  The two basis states both have the form of the phase-flip code, where each of the three qubits of that code has then been encoded in the bit-flip code.  A nested structure like this is called a *concatenated code*.  We can easily see how this code can correct a single bit-flip ($X$) error.  Each of the three triplets of qubits (123, 456, and 789) is a bit-flip code; one can detect and correct a single $X$ error without destroying the overall codeword state.  Less obviously, if a phase-flip ($Z$) error acts on one of the qubits of this code, it will flip the sign of that qubit's triplet from + to – or vice versa.  This moves the state to an orthogonal subspace, which can be detected and corrected without destroying the encoded state.

This code can therefore correct any single $X$ error *and* any single $Z$ error.  But this allows it to do even more.  Since $Y = iZX$, a $Y$ error can be thought of a single X error and a single $Z$ error acting on the same qubit, up to an irrelevant global phase.  So this code can correct any Pauli error acting on a single qubit.  However, even this is not the limit.  Note that *any* operator on a single qubit can be written as a linear combination $O = aI + bX + cY + dZ$ for some complex numbers $a, b, c, d$.  This operator acting on one qubit of the encoded state $|\psi_L\rangle = \alpha|0_L\rangle + \beta|1_L\rangle$ will produce a superposition of 4 orthogonal states:  the original state with no error; the state with a single $X$ error; the state with a single $Z$ error; and the state with both an $X$ and a $Z$ error.  The error correction procedure will project the codeword into one of these states, and then apply a correction that will return it to the original state $|\psi_L\rangle$.

This is one of the most important properties of QECCs, which makes them capable of handling more than idealized error models:  if a QECC has a correctable error set $\mathcal{E} = \{E_0, E_1, \ldots, E_{N-1}\}$ then it can also correct *any linear combination of these errors*, $E = a_0 E_0 + \cdots + a_{N-1} E_{N-1}$.  Thus, QECCs can correct a *continuous* set of errors.

Because the codeword is nine qubits long, it gives a benefit only for a lower value of the error probability per qubit than the 3-qubit repetition code.  If the error probability per qubit is $p$, then one will be better off encoding in the Shor code if $(1-p)^9 + 9p(1-p)^8 < 1 - p$, which is true for $p \lesssim 0.0323$.

*The Steane code.* Shor constructed a code that could correct an arbitrary single-qubit error by concatenating a code for correcting *X* errors with a code for correcting *Z* errors. Andrew Steane, working independently, showed that it was possible to do this in a code without this concatenated structure (Steane, 1996a). Steane's construction encodes a single logical qubit state into a 7-qubit codeword with the following basis vectors:

$$|0_L\rangle = \frac{1}{\sqrt{8}}(|0000000\rangle + |1010101\rangle + |0110011\rangle + |1100110\rangle + |0001111\rangle + |1011010\rangle$$
$$+ |0111100\rangle + |1101001\rangle),$$
$$|1_L\rangle = \frac{1}{\sqrt{8}}(|1111111\rangle + |0101010\rangle + |1001100\rangle + |0011001\rangle + |1110000\rangle + |0100101\rangle$$
$$+ |1000011\rangle + |0010110\rangle).$$

By the same argument as in the Shor code above, this code can also correct an arbitrary error on a single qubit. In this case, it takes considerable effort to verify from the codewords that any single *X* error moves the state $|\psi_L\rangle = \alpha|0_L\rangle + \beta|1_L\rangle$ to an orthogonal subspace, as does any single *Z* error and any combination of a single *X* error and a single *Z* error. Since this codeword is only seven qubits long, it shows a benefit for modestly higher values of the error probability per qubit *p*: $(1-p)^7 + 7p(1-p)^6 < 1 - p \Rightarrow p \lesssim 0.0579$.

*The error correction condition.* Knill and Laflamme (Knill and Laflamme, 1997) and independently Bennett, DiVincenzo, Smolin and Wootters (Bennett et al., 1996) found a very general condition for a set of errors to be correctable by a given quantum code. Let $\mathcal{E} = \{E_0, E_1, \ldots, E_{N-1}\}$ be some arbitrary set of error operators, and let *P* be a projector onto the code space. The code specified by *P* can correct the set of errors $\mathcal{E}$ if they satisfy the condition $PE_i^\dagger E_j P = \alpha_{ij} P$ for all *i* and *j*, where $\{\alpha_{ij}\}$ is a set of complex numbers that form a Hermitian matrix. The proof relies on diagonalizing this matrix and using that similarity transformation to construct a new set of errors that are linear combinations of the original set, and which map the state into orthogonal error spaces.

**Stabilizer Codes**

The two examples of the Shor and Steane codes (Shor, 1995; Steane, 1996a) show that powerful QECCs can be constructed that can encode general states and correct arbitrary errors on some number of qubits. But it is clear that better methods are needed for finding and describing these codes and their error-correction methods. Listing the code-space basis states, as done in the examples considered so far, rapidly becomes unwieldy, since the dimension of the Hilbert space grows exponentially with the number of qubits. Fortunately, a powerful formalism exists that can describe a large set of practical QECCs and their encoding and correction procedures in a very compact form. These are the *stabilizer codes* (Calderbank and Shor, 1996; Calderbank et al., 1997; Gottesman, 1996; Gottesman, 1997; Shor and Laflamme, 1997; Steane, 1996b). These codes use an error-correcting structure based on classical linear codes.

*Classical linear codes.* Suppose that we wish to encode *k* classical bits into a binary codeword that is an *n*-bit string. We can think of the *k* classical bits as being a *vector* **v** in a *k*-dimensional binary vector space. We encode that into a codeword which is a vector $\mathbf{v_L}$ in an $n$-dimensional binary space. This is a *linear code* if there is an $n \times k$ full-rank binary matrix $\mathbf{G}$ such that $\mathbf{v_L} = \mathbf{Gv}$. This matrix $\mathbf{G}$ is called the *generator matrix* of the code. Any valid codeword of the codes is a linear combination of the columns of $\mathbf{G}$, so the code forms a $k$-dimensional subspace of the $n$-dimensional binary space $\mathbb{Z}_2^n$.

For a linear code, we describe errors on the codeword by an $n$-dimensional binary error vector $\mathbf{e}$: $\mathbf{v_L} \rightarrow \mathbf{v_L} + \mathbf{e}$. In binary arithmetic, adding a 1 to a bit flips it from 0 to 1 or 1 to 0; so each element of $\mathbf{e}$ that is 1 represents a bit flip. (This model of errors is called *additive noise*.) To detect and correct errors, we define a second $(n-k) \times n$ full-rank binary matrix $\mathbf{H}$ called the *parity-check matrix*, which satisfies the equation $\mathbf{HG} = \mathbf{0}$. If we multiply an erroneous codeword by $\mathbf{H}$, we will get $\mathbf{H}(\mathbf{v_L} + \mathbf{e}) = \mathbf{Hv_L} + \mathbf{He} = \mathbf{0} + \mathbf{He} = \mathbf{He}$. The code should be designed so that the *error syndrome* (or *parity check*) $\mathbf{He}$ takes distinct, nonzero values for all of the most likely errors $\mathbf{e}$. From the error syndrome one can deduce which bits of the codeword have been flipped and flip them back to correct the error. For a small code, this diagnosis can be done by finding the error syndrome in a look-up table; for larger codes, this rapidly becomes impractical, and some form of *decoding algorithm* is necessary. In general, decoding for an arbitrary linear code is a computationally hard problem. But some codes have structure that allows efficient decoding.

Since $\mathbf{H}$ and $\mathbf{G}$ must satisfy $\mathbf{HG} = \mathbf{0}$, we can specify a linear code by giving either its generator matrix or its parity-check matrix. For the purposes of QEC, it is most convenient to use the parity-check matrix. The classical 3-bit repetition code is a simple linear code, where the parity-check matrix is

$$\mathbf{H} = \begin{pmatrix} 1 & 1 & 0 \\ 0 & 1 & 1 \end{pmatrix}.$$

Looking at the rows of this matrix one can see that their structure is echoed by the stabilizer generators *ZZI* and *IZZ* of the quantum bit-flip code, and *XXI* and *IXX* of the phase-flip code.

There are much more sophisticated linear codes than the repetition code, that encode larger numbers of bits and/or correct more errors. A well-known example is the 7-bit Hamming code:

$$\mathbf{H} = \begin{pmatrix} 0 & 0 & 0 & 1 & 1 & 1 & 1 \\ 0 & 1 & 1 & 0 & 0 & 1 & 1 \\ 1 & 0 & 1 & 0 & 1 & 0 & 1 \end{pmatrix}.$$

This classical linear code is closely related to the 7-qubit Steane code (as can be seen in the subsection on Calderbank-Shor-Steane (CSS) codes). A classical linear code is often described in terms of 3 parameters $[n, k, d]$, where $n$ is the number of physical bits (or length) of the code, $k$ is the number of logical (or encoded) bits, and $d$ is the minimum distance of the code—that is, the minimum number of 1s in any valid codeword other than the zero vector, or the minimum number of bits that must be flipped to transform one valid codeword into another. A code with minimum distance $d$ can correct any error that flips fewer than $d/2$ bits. The Hamming code has parameters [7,4,3]. These parameters all have equivalents for QECCs.

*Stabilizer groups and their generators.* Consider a subgroup $S$ of the Pauli group $\mathcal{G}_n$ on $n$ qubits with the following two properties: (1) the subgroup is Abelian (i.e., all operators in the subgroup commute); and (2) the subgroup does not contain the element -$I$. Then there is a subspace $C$ of the $n$-qubit Hilbert space $\mathcal{H}^{\otimes n}$ whose vectors are simultaneous +1 eigenvectors of all the elements of $S$. We call $S$ a *stabilizer group*, and $C$ is the code space corresponding to that stabilizer group.

We could specify $S$ by listing all of its elements. For small enough $n$ this can be possible; for large $n$, a typical stabilizer group has an exponentially large number of elements. However, we can specify $S$ much more compactly by listing a set of *stabilizer generators*. It is not hard to show that any stabilizer group on $n$ qubits must have $2^r$ elements, where $r$ is an integer between 0 and $n$. Every element in this group can be generated by a set of at least $r$ properly-chosen elements of $S$.

Consider a simple example. The bit-flip code is a stabilizer code, whose stabilizer group is given by $S = \{III, ZZI, IZZ, ZIZ\}$. We can choose two non-identity elements to be the stabilizer generators; for instance, $g_1 = ZZI$ and $g_2 = IZZ$. Then the 4 elements of the stabilizer are given by products of the generators: $(g_1)^i (g_2)^j$, where $i, j = 0$ or 1. Raising an operator to the zero power yields the identity; raising an operator to the first power is the operator itself. For the bit-flip code the four elements of the stabilizer $S$ are:

$$\begin{aligned} III &= (g_1)^0 (g_2)^0 = (III)(III), \\ ZZI &= (g_1)^1 (g_2)^0 = (ZZI)(III), \\ IZZ &= (g_1)^0 (g_2)^1 = (III)(IZZ), \\ ZIZ &= (g_1)^1 (g_2)^1 = (ZZI)(IZZ), \end{aligned}$$

where in the last expression we used the fact that $Z^2 = I$. This example generalizes for any stabilizer code. It is important to note that the elements of a stabilizer group are always Hermitian operators with eigenvalues ±1, and therefore can be measured as observables. Similarly, the phase-flip code has generators *XXI* and *IXX*.

*Error syndromes and error correction.* To understand how stabilizer codes detect and correct errors it is helpful to assume that the set of errors also consists of operators from the Pauli group $\mathcal{G}_n$. As we have seen in the examples of the Shor and Steane codes, this does not necessarily limit the types of errors that these codes can correct, since they will also be able to correct linear combinations of Pauli operators.

An important property of the Pauli group is that *any two Pauli operators either commute or anticommute*. For example, the Pauli operators *ZZI* and *XXX* commute (because anticommuting *X*s and *Z*s overlap at an even number of locations), while *ZIZ* and *YII* anticommute (because anticommuting *Y*s and *Z*s overlap at an odd number of locations).

Based on this property we can see how a stabilizer code detects and corrects errors. A valid codeword will be a +1 eigenvector of all the stabilizer generators. Suppose that an error operator $E$ (which is also an element of the Pauli group) multiplies the state. It will anticommute with some of the stabilizer generators, and commute with others. Multiplying by $E$ changes the codeword to a new eigenstate of the stabilizer generators, where the eigenvalue is still +1 for all the generators that commute with $E$, but is -1 for those generators that anticommute with $E$. This new eigenstate will always be orthogonal to the original codeword unless the error operator commutes with all the stabilizer generators.

Let's see how this works for the bit-flip code. It has generators $g_1 = ZZI$ and $g_2 = IZZ$. The three weight-one errors are $E_1 = XII, E_2 = IXI, E_3 = IIX$. We can see that $E_1$ anticommutes with $g_1$ and commutes with $g_2$; $E_2$ anticommutes with both $g_1$ and $g_2$; and $E_3$ anticommutes with $g_2$ and commutes with $g_1$. Since $g_1$ and $g_2$ are commuting observables, we can measure them to diagnose which error happened (or no error). The measured values ±1 for each generator are the error syndrome. Since Pauli operators are unitary and square to the identity, we can then undo the effects of the error by applying the appropriate Pauli operator again.

This is the general prescription for correcting errors with a stabilizer code. One measures the values ±1 of the stabilizer generators; from the resulting error syndrome, one deduces which error occurred, and applies the inverse (which for Pauli operators means just applying the error again). If the true error operator was actually a linear combination of Pauli operators, measuring the stabilizer generators will project the state into a joint eigenspace, and one proceeds exactly as if the error had been a Pauli operator. Just like linear codes, for a small stabilizer code one can use a look-up table of error syndromes; for a larger code, a decoding algorithm is needed.

Also like a classical linear code, a stabilizer code has three parameters, generally written $[[n, k, d]]$, where $n$ is the number of physical qubits of the code and $k$ is the number of logical qubits encoded (meaning that the code space has dimension $2^k$). The number of stabilizer generators is $r = n - k$. The third parameter $d$ is the *minimum weight* of any Pauli operator (other than the identity) that commutes with all the stabilizer generators. (The weight of a Pauli operator is the number of operators in the tensor product that are not the identity.) If a stabilizer code has distance $d$, it can correct an *arbitrary* error of weight $w \leq (d-1)/2$ (Lidar and Brun, 2013). If errors act independently on the qubits, and the error probability per qubit $p$ is not too large, then with high probability any errors that occur will be of low weight. The Shor code is $[[9,1,3]]$; the Steane code is $[[7,1,3]]$; and the bit-flip and phase-flip codes are both $[[3,1,1]]$. (They have distance 1 because they cannot correct arbitrary errors.)

*Calderbank-Shor-Steane (CSS) codes.* It turns out that the Shor and Steane codes are also stabilizer codes. With some work, one can identify a set of stabilizer generators for each of them. For the Shor code, a set of stabilizer generators is:

$$g_1 = Z_1Z_2, g_2 = Z_2Z_3, g_3 = Z_4Z_5, g_4 = Z_5Z_6, g_5 = Z_7Z_8, g_6 = Z_8Z_9,$$
$$g_7 = XXXXXXIII, g_8 = IIIXXXXXX.$$

Note the asymmetry in form between the generators involving $Z$ operators and those involving $X$ operators. This is because of the concatenated structure of the code. By contrast, we can find a set of stabilizer generators for the Steane code that are highly symmetric:

$$g_1 = IIIXXXX, \quad g_4 = IIIZZZZ,$$
$$g_2 = IXXIIXX, \quad g_5 = IZZIIZZ,$$

$$g_3 = XIXIXIX, \qquad g_6 = ZIZIZIZ.$$

Here, the generators involving *X* operators (which are used in detecting and correcting *Z* errors) and the generators involving *Z* operators (which are used in detecting and correcting *X* errors) have exactly the same form; moreover, the pattern of *I*s and *X*s (or *Z*s) exactly matches the pattern of 0s and 1s in the three rows of the parity-check matrix for the Hamming code.

Both the Shor and Steane codes have sets of stabilizer generators where one subset involves only *I*s and *X*s and the other involes only *I*s and *Z*s. Codes with this structure are called *Calderbank-Shor-Steane* (CSS) codes (Calderbank and Shor, 1996; Shor, 1995; Steane, 1996a; Steane, 1996b). The QECC can be thought of as the intersection of two classical linear codes, one in the *Z* basis (which can correct *X* errors) and one in the *X* basis (which can correct *Z* errors). In some cases—like the Steane code—these two linear codes are the same (the Hamming code in this case). In other cases—like the Shor code—the two linear codes are different. It is possible to separately correct *X* and *Z* errors just as one would for a classical code (though that is generally not optimal, since *X* and *Z* errors may be correlated). The size of the intersection determines the number of logical qubits; the Hamming code encodes four bits, but the Steane code encodes only one qubit, because the intersection is only two-dimensional.

However, there is a very important constraint on how codes can be combined in this way to make a CSS code, because the final set of stabilizer generators must all commute. This constraint turns out to be equivalent to an orthogonality condition for the original codes: if the code for correcting bit-flips has parity-check matrix $\mathbf{H}_Z$ and the code for correcting phase-flips has parity-check matrix $\mathbf{H}_X$, then they must satisfy the binary matrix equation $\mathbf{H}_X \mathbf{H}_Z^T = \mathbf{0}$. This means that for a code like the Steane code that uses the same classical code for both *X* and *Z*, the parity-check matrix must be *self-orthogonal*: $\mathbf{H}\mathbf{H}^T = \mathbf{0}$.

CSS codes are not the only way to construct stabilizer codes from classical linear codes. A more general method gives the Calderbank-Rains-Shor-Steane (CRSS) codes (Calderbank et al., 1997). The parity-check matrices in this case obey a more general self-orthogonality condition analogous to the CSS case.

*Degenerate codes*. Stabilizer codes inherit many of the properties of classical linear codes, but there are certain properties unique to quantum codes. These arise in part because, while the codes are modeled on codes with an additive structure, the noise is actually multiplicative. One such property is known as *degeneracy*. Consider, again, the Shor code. The errors $Z_1, Z_2, Z_3$ all transform the code space in *exactly* the same way. They all have the same error syndrome and can be corrected by the same unitary correction operator. This degeneracy arises because these operators differ by an element of the stabilizer group: $Z_1 = Z_2(Z_1 Z_2) = Z_3(Z_1 Z_3)$. Since the operators $Z_1 Z_2$ and $Z_1 Z_3$ stabilize the codeword, these three errors have identical effects. These errors cannot be distinguished from each other by the code, but they are all corrected by applying $Z_1$. This can happen with *any* stabilizer code: an error *E* has the same effect as *ES* where *S* is any element of the stabilizer group.

We say a given code is *degenerate* if its correctable error set contains degenerate errors. For an $[[n, k, d]]$ stabilizer code, the correctable error set is generally taken to include all Pauli errors with weight $w \leq (d-1)/2$ (which of course implies the ability to correct *arbitrary* errors of that weight and lower). For example, both the Shor code and Steane code have minimal distance $d = 3$, which means they can correct all errors of weight 1; but the Shor code is degenerate, while the Steane code is not.

*Logical operators.* An $[[n, k, d]]$ stabilizer code has $n - k$ independent stabilizer generators. But there are operators in the Pauli group $\mathcal{G}_n$ that are not elements of the stabilizer group but commute with every stabilizer generator. These are called *logical* operators, because they act directly on the encoded logical qubits. Such operators could represent encoded quantum gates; but they can also represent undetectable errors. Any code with $k > 0$ has logical operators.

It is common to write down a set of canonical logical operators for a code, comprising $k$ anticommuting pairs, one for each encoded logical qubit. For example, the bit-flip code has logical operators *XXX* and *ZII*, which act as Pauli operators *X* and *Z* on the encoded qubit. For the phase-flip code the equivalent operators are *XII* and *ZZZ*, and for the Steane code they are *XXXXXXX* and *ZZZZZZZ*. It's important to remember, however, that one can multiply a logical operator by any element of the stabilizer group without changing its action on the codewords. So for the Steane code, the operators *XXXIIII* and *ZZZIIII* are an equivalent pair of logical operators. The lowest-weight logical operator (aside from the identity) has a weight equal to *d*.

*The 5-qubit code and general stabilizer codes.* All of the codes examined so far have been CSS codes, which might give the impression that most stabilizer codes are CSS codes. This is certainly not the case (though CSS codes are widely used in fault-tolerant quantum computation). The first non-CSS code discovered was the 5-qubit code, discovered independently by Bennett, DiVincenzo, Smolin and Wootters and by Laflamme, Miquel, Paz and Zurek (Bennett et al., 1996; Laflamme et al., 1997). This code has parameters [[5,1,3]]; there are several variations of this code, but they are all equivalent in their properties. One version of this code has stabilizer generators

$$\begin{aligned} g_1 &= XZZXI, & g_3 &= XIXZZ, \\ g_2 &= IXZZX, & g_4 &= ZXIXZ. \end{aligned}$$

None of these stabilizer generators involves only *X*s or only *Z*s, and it is not hard to show that there is no set of generators for this code that does. Codes like this cannot be interpreted as the intersection of a code for *X* errors and a code for *Z* errors.

This code encodes a single logical qubit in five physical qubits and can protect against an arbitrary error on any one qubit. It is shorter than the Steane code, and like the Steane code it is not degenerate. In fact, one can show that no quantum code that can protect against an arbitrary single qubit error can be shorter than 5 qubits. If a code has *n* physical qubits, it must be able to distinguish $3n + 1$ distinct error syndromes (*X*, *Y* and *Z* on each physical qubit, plus the identity). If it has $n - k$ stabilizer generators there are $2^{n-k}$ distinct error syndromes. So we must have $3n + 1 \leq 2^{n-k}$. For *k* = 1 the smallest *n* that satisfies this inequality is *n* = 5, where the two sides are equal. For this reason, the 5-qubit code is sometimes called a "perfect" code.

*Encoding and decoding circuits.* This discussion of quantum codes and error correction has been rather abstract, in terms of encoding into subspaces and measurements of observables. How would this be done in practice? In quantum information science, one generally decomposes unitary transformations and measurements into *quantum circuits*, which are sequences of *quantum gates*: unitary transformations that act on only one or two qubits at a time. There are standard sets of quantum gates that are widely used. On one qubit, common gates include the Hadamard (*H*), Phase (*S*) and π/8 (*T*) gates,

$$H = \frac{1}{\sqrt{2}}\begin{pmatrix} 1 & 1 \\ 1 & -1 \end{pmatrix}, \quad S = \begin{pmatrix} 1 & 0 \\ 0 & i \end{pmatrix} = \sqrt{Z}, \quad T = \begin{pmatrix} 1 & 0 \\ 0 & e^{i\pi/4} \end{pmatrix} = \sqrt{S},$$

as well as the usual Pauli operators (Nielsen and Chuang, 2000). On two qubits the most common gates are the controlled-NOT (CNOT) and controlled-phase (CZ) gates. Using these simple gates, we can write down encoding circuits for the bit-flip and phase-flip codes (see Fig. 1). Extracting the error syndromes can also be done by a quantum circuit (see Fig. 2). These error syndromes specify the correction to be done, if any.

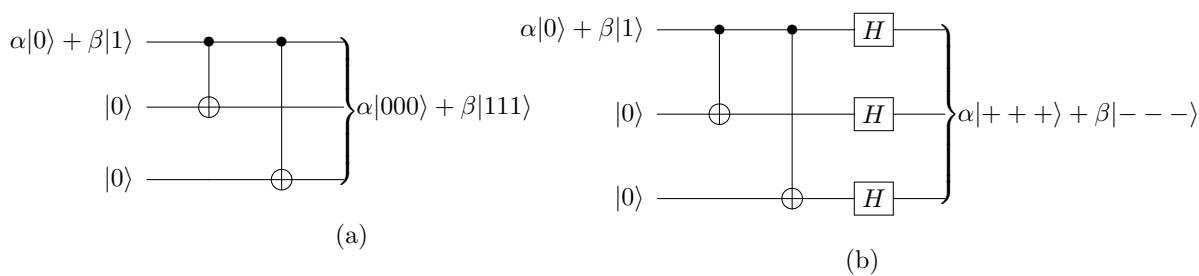

Fig. 1: (a) The encoding circuit for the 3-qubit bit-flip code, involving 2 CNOT gates. (b) The encoding circuit for the 3-qubit phase-flip code. Encoding involves 2 CNOT gates and 3 Hadamard gates.

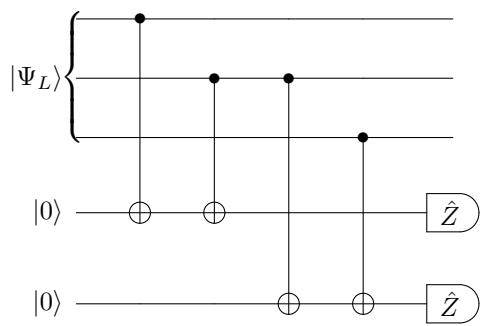

Fig. 2: The syndrome extraction circuit for the 3-qubit bit-flip code. Two CNOTs put each parity check in one of the two extra qubits at the bottom (called *ancillary qubits* or *ancillas*), which is then measured in the standard *Z* basis. The outcome of these measurements is an error syndrome, which tells what correction should be applied (if any). For syndrome (+1,+1), no correction is done; for (+1,-1), *X* is applied to qubit 3; for (-1,+1), *X* is applied to qubit 1; and for (-1,-1), *X* is applied to qubit 2.

An interesting property of stabilizer codes is that their encoding and syndrome-reading circuits can always be written using just three kinds of quantum gates: the CNOT, the *H*, and the *S*. These three gates generate (up to a global phase) a subgroup of the unitary group called the *Clifford group*. This is the set of unitaries *U* that preserve the Pauli group under similarity transformations: $\forall g \in \mathcal{G}_n, UgU^\dagger \in \mathcal{G}_n$. So one can think of encoding and decoding circuits as transforming one stabilizer group into another. Another interesting property is that, given the ability to do arbitrary Clifford unitaries, plus any *one* unitary outside the Clifford group, one can generate any *arbitrary* unitary transformation on *n* qubits. Such a gate set is *universal*. But encoding and decoding can be done just with the Clifford gates (Gottesman, 1997).

**Fault-Tolerance and Error Correction for Quantum Computation**

Up to this point we have only considered the question of protecting *static* quantum information from noise. This is a typical model of quantum communication or storage: it is assumed that the encoding and correction/decoding of the quantum information are error-free, and that decoherence happens only during transmission through a noisy channel. Ignoring the errors during encoding and decoding is an idealization, but it is reasonable to separate the effects of errors due to imperfect quantum circuits from the unavoidable errors in passing through the channel.

This separation does *not* make sense when we consider quantum computation. Here we are not only storing information but processing it. If left unchecked, errors can accumulate and spread during processing until they are uncorrectable, which would restrict the size of computations that could be done. To combat this, repeated error corrections are required, so one must consider the effects of errors during the correction process itself. *Fault-tolerant quantum computation* (FTQC) is the set of principles that allows the use of repeated error correction during a long quantum computation without introducing more errors than are corrected (Aharonov, 1999; Aharonov et al., 1996; Aharonov and Ben-Or, 1997; Aharonov and Ben-Or, 1999; Aliferis, Gottesman and Preskill, 2006; DiVincenzo and Shor, 1996; Gottesman, 1997; Gottesman, 1998; Kitaev, 1997a; Kitaev, 1997b; Knill, Laflamme and Zurek, 1998a; Knill, Laflamme and Zurek, 1998b; Knill, Laflamme and Viola, 2000; Preskill, 1997; Preskill, 1998a; Preskill, 1998b; Reichardt, 2005a; Shor, 1996).

*Principles of fault tolerance*. FTQC is a very large topic in itself, so we can only touch on some of the basic ideas. Here are a few of the guiding principles:

1. Never decode the quantum information. All operations must be done on the encoded quantum data.
2. Quantum circuits acting on encoded data should be robust against errors. The circuits should not cause a correctable error to spread until it is an uncorrectable error.
3. The encoded information should be corrected periodically, to catch and remove errors before they accumulate.
4. Error correction circuits also should not spread errors.

5. It is impossible to ever remove all errors; but any residual errors should be correctable, so they can be caught and removed in the next error-correction step.

The requirements of fault tolerance are quite stringent, and not every QECC can meet them. One QECC that looks more powerful than another "on paper" may be less suitable for FTQC. For example, it is much harder to do encoded operations on the 5-qubit "perfect" code than on the 7-qubit Steane code. CSS codes are widely used in FTQC because their additional structure makes it easier to design fault-tolerant circuits for them.

*Encoded gates*. FTQC usually starts with a QECC (or a family of QECCs) and designs fault-tolerant circuits for a universal set of encoded gates. As mentioned at the end of the subsection on encoding and decoding circuits, the ability to do CNOTs, Hadamard gates, Phase gates and any one non-Clifford gate implies universality. A very common approach to FTQC is to use a code that allows efficient encoded Clifford gates, and then use a more difficult technique to do a non-Clifford gate (most often the π/8 gate). For example, the Steane code allows all Clifford gates to be done *transversally*, that is, by applying a gate to each qubit separately.

Transversal gates are particularly useful for fault-tolerance because they do not spread errors from one qubit to another. In the case of the transversal CNOT, a single-qubit error on one codeword can spread to a different codeword, but it cannot spread to a second qubit on the same codeword. Various approaches are available to do non-Clifford gates fault-tolerantly; one of the most common is to prepare a special state, called a *magic state*, which can be input into a circuit using only Clifford gates to effectively produce a non-Clifford gate. Low-error encoded versions of these states are prepared by a process called *magic state distillation* (Bravyi and Haah, 2012; Bravyi and Kitaev 2005; Haah et al., 2017; Knill, 2004a; Knill, 2004b; Reichardt, 2005b).

*Threshold theorems*. A key element underlying FTQC is a set of results called *threshold theorems*. These theorems prove that if the physical rate of errors is below a certain value, called the *error threshold*, then it is possible to perform a quantum computation of arbitrary size (Aharonov, 1999; Aharonov et al., 1996; Aharonov and Ben-Or, 1997; Aharonov and Ben-Or, 1999; Aliferis, Gottesman and Preskill, 2006; Gottesman, 1997; Knill, Laflamme and Zurek, 1998a; Knill, Laflamme and Zurek, 1998b; Preskill, 1998b; Reichardt, 2005a; Shor, 1996). The additional overhead for all the extra error correction and fault-tolerant design (over an ideal, error-free quantum computer) scales like a polynomial in the log of the size of the ideal circuit (that is, the total number of gates). This scaling grows quite slowly, so in principle one can scale up to very large quantum computations with only relatively modest overhead.

The first threshold theorems used the idea of a concatenated code. Each qubit of the ideal circuit is replaced by a code word (for example, in the Steane code), and each gate is replaced by a circuit for an encoded gate. After each encoded gate, an error correction step is performed. If the success probability is still too low, one iterates this encoding process as many times as necessary to bring the success probability up to a desired value.

A simple back-of-the-envelope argument shows why this can work. Suppose that the error rate per gate is $p \ll 1$, and the ideal circuit has $N$ gates, so the success probability scales like $(1-p)^N \approx e^{-pN}$. Suppose we encode each qubit in a code that can correct any single-qubit error. Then the probability of an uncorrectable error per *encoded* gate becomes roughly $Cp^2$, where $C$ is a constant representing the increased size of the circuit. If we iterate this process $k$ times ($k$ levels of concatenation) then the rate of uncorrectable errors becomes $(Cp)^{2^k}/C$. So if $p < 1/C$, then the rate of uncorrectable errors goes down *doubly exponentially*, while the size of the circuit grows only *singly exponentially* (roughly like $C^{k/2}$).

Underlying these theorems is a set of assumptions that might or might not hold in realistic quantum computers. Typically, these assumptions resemble the following:

1. The quantum computer allows parallel operations (in particular, error correction can be done in parallel throughout the computer).
2. Errors are not strongly correlated across the qubits in the computer (the probability of a high-weight error should fall off like a binomial distribution).
3. Errors on quantum gates affect the qubits taking part in the gates, but not other unrelated qubits. Two-qubit gates can be done between any pair of qubits.
4. Qubit measurements can be done quickly, and their error rates are not much higher than the error rates for quantum gates.
5. Memory errors occur at a rate less than the rate for quantum gates.
6. Information about the error syndromes for the quantum codes can be processed (classically) quickly and without errors.

This set of assumptions can be relaxed in various ways while still producing a threshold theorem, but this can adversely affect the size of the threshold. The threshold depends strongly on the code being used, the particular error model, and the choice of fault-tolerant methods. Early threshold theorems estimated error thresholds of $10^{-6}$ errors per gate or smaller (Aliferis, Gottesman and Preskill, 2006), a very difficult goal to meet experimentally. Improvements in both fault-tolerant methods and proofs have gradually raised these thresholds, so that concatenated codes can now have thresholds of $10^{-4}$ or higher (Chamberland et al., 2016). It is also possible (up to a point) to trade off a higher threshold for larger overhead (Knill, 2004a; Knill, 2004b). However, the exponential growth of the circuit size is a large technological barrier, even if the asymptotic scaling rate is reasonable.

More recent work has improved matters considerably by looking at different classes of codes: the *topological* codes (Bravyi and Kitaev, 1998b; Freedman and Meyer, 1998; Kitaev, 1997a; Kitaev, 2003; Preskill, 1997), especially the surface code (Bravyi, Suchara and Vargo, 2014; Fowler, Stephens and Groszkowski, 2009; Fowler, Wang and Hollenberg, 2011; Fowler et al., 2012), but also color codes, among others (Bombin, 2015; Brown, Nickerson and Browne, 2016; Kubica and Beverland, 2015; Landahl, Anderson and Rice, 2011). These encode quantum information in a (usually two-dimensional) lattice of qubits and are designed to mimic the topologically-protected properties of 2D quantum field theories with non-Abelian anyons (Kitaev, 2003). These are generally CSS codes, but the stabilizer generators are all local operators on the lattice with bounded weight, making them much easier to measure. (See Fig. 3.)

Encoded operations are done by a combination of *code deformation* (such as braiding lattice defects around each other) and magic state distillation. Threshold theorems have been proven for these codes that suggest an error threshold of $10^{-3}$ or so, but numerical evidence suggests that the practical threshold may be even higher than that (on the order of 0.5%) (Stephens, 2014). Moreover, the distances of these codes can be scaled up linearly, rather than in the exponential leaps of concatenated codes, which makes their overhead scale more nicely. These properties have made surface codes and other topological codes the subject of intense theoretical and numerical research, as well as of experimental plans (Barends et al., 2014).

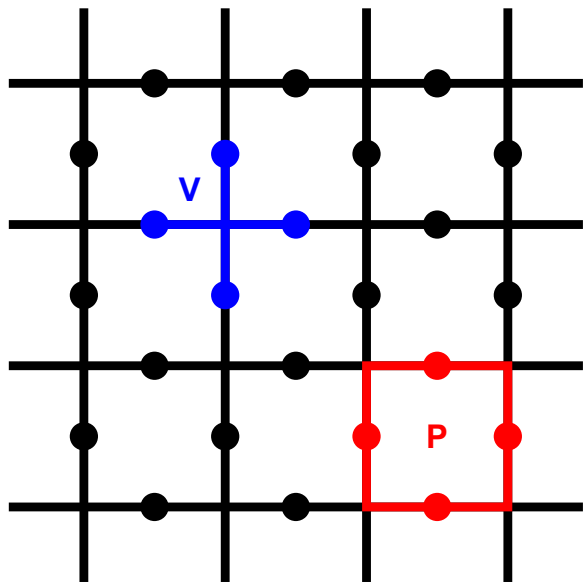

Fig. 3: The qubit lattice for the surface code. The qubits are located on the edges; the stabilizers comprise 2 types of local operators. *X* operators are located around *plaquettes* (P in the figure), while *Z* operators are located around *vertices* (V in the figure).

With the recent development of small, noisy quantum computers, other approaches (Brun et al., 2015; Chao and Reichardt, 2018a; Chao and Reichardt, 2018b; Lai, Zheng and Brun, 2017; Reichardt, 2018; Steane, 1999; Steane, 2003; Zheng, Lai and Brun, 2018) have also been explored, which might reduce the overhead enough for FTQC to be possible in near-term experiments. FTQC is a very active field, with constant progress being made; it is fair to say that we do not know what methods practical quantum computers of the future will use.

**Variations and Generalizations of Quantum Error Correction**

*Other kinds of quantum noise.* In practice, the physical systems used as qubits are often not truly two-level systems. For instance, in quantum communication it is common to use the polarization states of a single photon to represent a qubit, but such photonic channels are subject to photon loss, in which the system goes to the vacuum state. Another common qubit uses two hyperfine levels of a trapped ion as a qubit; it is possible for the ion to make an unintended transition to an excited state. Such errors (where the system leaves the qubit subspace spanned by $\{|0\rangle, |1\rangle\}$) are generically called *leakage* errors. It is quite possible to design QECCs to deal with leakage errors as well as the kind of qubit errors considered in this article. Indeed, leakage errors are sometimes easier to correct; a measurement can reveal that a qubit is not in the usual subspace, and then treat that as an *erasure* error (which, being at a known location, is easier to correct).

It is also commonly assumed that the errors are *Markovian*, so the qubits do not interact repeatedly with the same environment degrees of freedom. Many realistic systems, however, are *non-Markovian*. This is not always a serious problem, but it certainly makes it harder to model the error process.

*Convolutional codes*. The QECCs described in this article are all *block codes*: that is, they encode a fixed number of logical qubits $k$ into a fixed number of physical qubits $n$. There is another type of code, called a *convolutional* code, which works differently: the logical qubits arrive in a steady stream, and as they arrive, they are encoded and transmitted. The ratio of logical to physical qubits is fixed, but the length of the codeword can vary tremendously. Classical convolutional codes are widely used in communication; quantum convolutional codes (Forney, Grassl and Guha, 2007; Grassl and Rötteler, 2006; Ollivier and Tillich, 2003) have been explored for similar use in quantum communication.

*Generalizations and extensions of stabilizer codes*. The QECCs described in this article have, for the most part, been qubit codes. But there has also been work on error-correcting codes for $d$-dimensional quantum systems, or *qudits* (Gottesman, 1999; Rains, 1999). The Pauli operators can be generalized in a number of ways to sets of $d \times d$ matrices, and connections made to linear codes over different finite fields.

For qubit codes, the basic stabilizer formalism described in this article can be generalized in a number of ways. One is in terms of *operator* or *subsystem* codes (Kribs, Laflamme and Poulin, 2005; Kribs et al., 2006; Kribs and Spekkens, 2006; Nielsen and Poulin, 2007; Poulin, 2005). One can think of these codes as including, in addition to the logical qubits that are protected from noise, some additional *gauge* qubits which are not protected. No information can be stored in these gauge qubits, but their presence can be used to reduce the complexity of encoded operations (Bacon and Casaccino, 2006).

Another extension is to *entanglement-assisted* codes (Bennett et al., 1996; Bowen, 2002; Brun, Devetak and Hsieh, 2006; Brun, Devetak and Hsieh, 2014), which can be used for quantum communication. These codes assume that the sender and receiver share some number of maximally entangled states prior to communication. This entanglement can dramatically boost the power of quantum codes, increasing their rate, or their ability to correct errors, or both. It can also relax the self-orthogonality constraint in constructing QECCs from classical linear codes. It is also possible to construct entanglement-assisted operator codes (Brun, Devetak and Hsieh, 2007; Hsieh, Devetak and Brun, 2007).

A larger class of codes—which include the stabilizer codes as a special case—are the *codeword stabilized* (CWS) codes (Cross et al., 2009; Van den Nest et al., 2004). These codes have received a limited amount of study; in some ways they can be more powerful than stabilizer codes, but general methods for constructing large CWS codes are not known, and nor are efficient methods for encoding and decoding them.

*Other methods of error mitigation*. QECCs and fault-tolerance are the main methods known to protect quantum computers from decoherence, but there are other methods that can make a significant improvement in some cases. *Decoherence-free subspaces* and *noiseless subsystems* (Bacon et al., 2000; Lidar, Chuang and Whaley, 1998; Zanardi, 2001; Zanardi and Rasetti, 1998) are encodings in which all errors are corrected *passively*: that is, stored information is immune to the effects of noise. Unfortunately, such subspaces (or subsystems) are not guaranteed to exist; they are generally only present when the noise process has major symmetries that can be exploited. *Dynamical decoupling* (DD) (Byrd and Lidar, 2002a; Duan and Guo, 1999; Khodjasteh and Lidar, 2005; Viola and Lloyd, 1998; Viola, Knill and Lloyd, 1999; Viola, Lloyd and Knill, 1999; Vitali and Tombesi, 1999; Zanardi, 1999), by contrast, is an active technique in which fast unitary pulses are applied to the quantum system in a regular pattern that causes the noise to average away to zero. DD has great advantages: it does not require measurements and feedback and can be applied to an entire set of qubits at once. But it also has limitations: it is only effective against non-Markovian noise, and it cannot increase the purity of a quantum state. However, it is possible to combine DD and QEC into a powerful hybrid approach to controlling errors in quantum systems (Byrd and Lidar, 2002b; Ng, Lidar and Preskill, 2011).

**Experimental Implementation of Quantum Error Correction**

While the theory of Quantum Error Correction has been worked out in considerable detail, the experimental challenges in realizing it are daunting. In encoding quantum data, one uses a redundant representation that will in general undergo more noise than the original data would have done if left unencoded. For instance, encoding a single qubit in a [[7,1,3]] Steane code might well increase the overall rate of errors sevenfold. The encoding circuit and the circuits for measuring error syndromes and applying corrections will all be subject to errors. Unless the intrinsic error rate for all these operations is low, the net effect of error correction will be worse than doing nothing. Fault tolerance is even more demanding, since it requires the processing as well as the storage of encoded data, and generally needs the preparation of high-quality ancillary states.

However, these difficulties also hold a promise: if error rates can be reduced to a sufficiently low level, then QEC can effectively make them as low as one desires. It will be possible to scale quantum computers to allow computations of unlimited size. That is the promise that supports and propels the entire field.

A number of experiments have been done to prove the principles of QEC. Many of these experiments applied artificial errors to a codeword and showed that error correction produced an improvement in the quality of the state (Chiaverini et al., 2004; Cory et al., 1998; Knill et al., 2001; Pittman et al., 2011; Reed et al., 2012). Others have demonstrated the necessary operations for QEC, but not shown an actual extension in the lifetime of the stored quantum state (Kelly et al., 2015; Schindler et al., 2011). As of this writing (in 2019), only one experiment has implemented QEC against the native noise in its system and shown a net gain over not using QEC at all (Ofek et al., 2016). This experiment encoded qubits as nonclassical states of a superconducting resonator with particular symmetry properties, and showed a moderately longer coherence lifetime than an unencoded qubit. This result, modest though it is, shows that experimental systems are approaching the point where the methods of QEC will become viable technologies.

**Conclusions and Open Questions**

Quantum error correction is the centerpiece of the current effort towards quantum information processing, both for quantum computers and quantum communication. Despite early naïve intuitions to the contrary, error correction turns out to be both possible and practical for quantum systems; by careful design of quantum error-correcting codes, it is possible to deduce and correct an error on a quantum system without gaining any information about (and hence disturbing) the protected quantum state. The most widely used and studied class of quantum codes are the stabilizer codes, which inherit much of their error-correcting structure from classical linear codes. Stabilizer codes are just one element of fault-tolerant quantum computation, which should allow quantum computations of arbitrary size to be done with high success probability, providing that the physical error rate is below an error threshold.

Quantum error correction is an extremely active field of research, and it is fair to say that we do not yet know the true limits of these techniques. Threshold theorems have been proven for a variety of codes and error models, but we really do not know how high a threshold can be achieved; numerical evidence suggests that an error threshold as high as a few percent might be possible. Moreover, many of the methods that have been studied have been aimed at proving asymptotic scaling results and may require highly unrealistic amounts of overhead (Fowler et al., 2012; Knill, 2004b; Lai et al, 2014; Reichardt, 2004; Steane, 1999b; Svore et al., 2005). As small quantum computers have started to become available, the focus is turning towards greatly reducing this overhead while maintaining error-correcting power (Brun et al., 2015; Chao and Reichardt, 2018a; Chao and Reichardt, 2018b; Reichardt 2018). In practice, we will only ever do computations of a finite size; ultimate scaling limits are of largely theoretical interest. A more practical question is this: for a given level of noise in a finite-sized quantum computer, how large a computation can be done with reasonable probability of success?

Prototype quantum computers have raised another important question: how well does the decoherence in real devices resemble the idealized assumptions underlying proofs of threshold theorems? There is strong evidence that current qubits are very far from having independent Markovian noise. Fortunately, recent theoretical and experimental work (Huang, Doherty and Flammia, 2019; Knill, 2005; Viola and Knill, 2005; Wallman and Emerson, 2016; Ware et al., 2018) also suggests that new fault-tolerant methods (such as *randomized Pauli frames*) may be able to transform quite general error models to resemble the Markovian Pauli error models commonly used in QEC research.

It is likely that the best codes and decoding algorithms for quantum computation have yet to be discovered. As larger and less noisy quantum computers become available, we will increasingly be able to test the performance of error correction ideas on real machines, and to build on the dramatic theoretical progress that we have already seen. In less than twenty-five years, quantum error correction has gone from a few scattered ideas, beset by skepticism and misunderstanding, to the large, vibrant field it is today. There is every reason to think that we are only at the beginning of what can be done.

**Further Reading**

For a broad overview of QEC, the book *Quantum Error Correction*, edited by Lidar and Brun, contains chapters on a wide range of topics in quantum error correction and error mitigation, contributed by top experts in the field (Lidar and Brun, 2013).

There is also a textbook, *Quantum Error Correction and Fault Tolerant Quantum Computing*, by Frank Gaitan, which presents the theory of QEC in a systematic fashion, with exercises, aimed at graduate students, advanced undergraduates, or anyone with a good technical background who wants to learn the field (Gaitan, 2008).

On a more narrowly-focused level, Daniel Gottesman's 1997 Ph.D. thesis, *Stabilizer Codes and Quantum Error Correction*, was a key contribution to the field of QEC, and is still one of the best introductions to stabilizer codes (Gottesman, 1997).

There is a recent review article on QEC and other methods of error prevention in *Reviews of Modern Physics*, "Protecting quantum information against environmental noise," by Suter and Alvarez (Suter and Alvarez, 2016).

There are several excellent textbooks on Quantum Information Science that include useful introductions to QEC. Still the most useful, nearly 20 years after it was written, is *Quantum Computation and Quantum Information*, by Michael A. Nielsen and Isaac L. Chuang (Nielsen and Chuang, 2000). While our understanding has advanced significantly since this book appeared, especially on experimental systems, as a presentation of the foundations of the field it is by far the best and most comprehensive.

An excellent, much shorter, introduction at the undergraduate level is *Quantum Computer Science: An Introduction*, by N. David Mermin (Mermin, 2007).

For those who want a more physics-centered approach to the subject, the textbook *Principles of Quantum Computation and Information*, by Benenti, Casati, Rossini and Strini is very good (Benenti et al., 2019). A somewhat eclectic, but very interesting, textbook is *Explorations in Quantum Computing*, by Colin P. Williams (Williams, 2011). For a more computer-science-based view, there is *Quantum Computing for Computer Scientists*, by Yanofsky and Mannucci (Yanofsky and Mannucci, 2008), and the interesting collection of musings in *Quantum Computing Since Democritus*, by Scott Aaronson (Aaronson, 2013).

*Quantum Information and Computation* by Lo, Popescu and Spiller, is an older book, but still interesting, and gives insight into the rapid development of the field in the 1990s (Lo, Popescu and Spiller, 1998).